\documentstyle[12pt,aasms4]{article}
\lefthead{Rosolowsky et al.}
\righthead{The Spectral Correlation Function}

\begin{document}
\title{The Spectral
Correlation Function --\\ A New Tool for Analyzing Spectral-Line Maps}

\author{Erik W. Rosolowsky\altaffilmark{1}, 
Alyssa A. Goodman\altaffilmark{2} \altaffilmark{3}, David
J. Wilner\altaffilmark{4} \altaffilmark{5} 
and Jonathan P. Williams\altaffilmark{6}}
\affil{Harvard-Smithsonian Center for Astrophysics, 
60 Garden Street, Cambridge, MA 02138.}
\altaffiltext{1}{erosolow@cfa.harvard.edu, Swarthmore College}
\altaffiltext{2}{agoodman@cfa.harvard.edu}
\altaffiltext{3}{National Science Foundation Young Investigator}
\altaffiltext{4}{dwilner@cfa.harvard.edu}
\altaffiltext{5}{Hubble Fellow}
\altaffiltext{6}{jpw@cfa.harvard.edu}
\vskip 2 in
\centerline{
Revised for {\it The Astrophysical Journal, Part I}, March 1999}

\begin{abstract}

The ``spectral correlation function'' analysis we introduce in this paper is a new
tool for analyzing spectral-line data cubes.  Our initial tests, carried out on a
suite of observed and simulated data cubes, indicate that the spectral correlation
function [SCF]  is likely to be a more discriminating statistic than other
statistical methods normally applied.  The SCF is a measure of similarity between
neighboring spectra in the data cube.  When the SCF is used  to compare a data cube
consisting of spectral-line observations of the ISM with a data cube derived from
MHD simulations of molecular clouds,  it can find differences that are not found by
other analyses.  The initial results presented here suggest that the inclusion of
self-gravity in numerical simulations is critical for reproducing the correlation
behavior of spectra in star-forming molecular clouds.

Subject headings: ISM: structure---line: profiles---(magnetohydrodynamics:)
MHD---methods: data analysis---molecular data---turbulence

\end{abstract}

\section{Introduction}

The ``spectral correlation function'' analysis we introduce in this short paper is
a new tool for analyzing spectral-line data cubes.  Owing to the recent advances in
receiver and computer technology, both observed and simulated cubes have been
growing in size.  Our ability to intuit their import, however, has not kept pace. 
Therefore, the need for statistical methods of analyzing these cubes has now become
acute.

Several methods for analyzing spectral-line data cubes have been proposed
and applied over the past fifteen years.  Many of the methods are
``successful'' in that they can describe a cube with far fewer bits than the
original data set contained.  The question in the study this paper
introduces can be phrased as ``just which bits describe the cube most
uniquely?"  In particular, we seek a method which produces easily understood results, but
preserves as much information as possible about {\it all} of the dimensions of a
position-position-velocity cube of intensity measurements.

Some previous statistical analyses do not explicitly make use of the
velocity dimension in analyzing spectral-line cubes.  For example, Gill and
Henriksen (1990) and Langer, Wilson, \& Anderson (1993) apply wavelet
analysis to position-position-intensity data, in order to represent the
physical distribution of material in a mathematically efficient way.
Houlahan and Scalo (1992) use structure-tree statistics on IRAS images to
analyze the hierarchical vs.  random nature of molecular clouds, ultimately
finding evidence for some of each.  Wiseman and Adams (1994) use
pseudometric methods on IRAS data to describe and rank cloud ``complexity."
Elmegreen and Falgarone (1996) analyze the clump mass spectrum of several
molecular clouds in order to determine a characteristic fractal dimension
for the star-forming interstellar medium.  Blitz and Williams (1997) find
evidence for a break in the column density distribution 
of material in clouds by
analyzing histograms of column density.

Other analyses preserve velocity information along with the spatial
information in analyzing the cubes.  At present, these kinds of analyses can
essentially be broken down into two groups.

In the first group, no transforms are taken, and spatial information is
preserved directly.  For example Williams, de Geus, \& Blitz (1994) use the
CLUMPFIND program, and Stutzki \& G\" usten (1990) use their GAUSSCLUMPS 
algorithm to identify ``clumps'' in position-position-velocity space.
Statistical analyses are made on the distributions of clump
properties (e.g.  the clump mass spectrum is calculated) to probe
the three dimensional structure of molecular clouds.

In the second group, transforms of one kind or another are performed, and
spatial information is preserved as ``scale'' rather than as ``position''
information.  The classic example of this kind of analysis involves
calculation of autocorrelation and structure functions.  Application of
these functions to molecular cloud data was first suggested by Scalo (1984)
and then applied to real data by Kleiner and Dickman (1985) and by Miesch
and Bally (1994).  Heyer and Schloerb (1994) have recently applied
Principal Components Analysis (PCA) to several data cubes.  This
method describes clouds as a sum of special functions in a 
manner mathematically similar to wavelet analysis.  
Most of these analyses have offered new
insights into cloud structure and kinematics.

Using this breakdown, the SCF falls
into the first group,\footnote{Strictly speaking, the SCF is in the first group when
applied for a fixed spatial resolution.  However, the SCF can be used as a tool
more like the autocorrelation function analyses mentioned in the second group, by
comparing runs with different spatial resolution.  An upcoming paper
(Padoan \& Goodman 1999) discusses the effects of varying the ratio of resolution
to map size on the SCF (see \S 3.4).} in that no transforms are performed and spatial
information is preserved directly. The SCF simply describes the similarities in shape,
size, and velocity offset among neighboring spectra in a data cube.  In originally
developing the SCF, our goal was to create a ``hard-to-fool'' statistic for use in
comparing data cubes calculated from simulations of the ISM with those of observed cubes.

The exact reproduction of an observed object in the ISM through simulation
is practically impossible so simulations need to be evaluated on their
ability to reproduce more general properties of the ISM like appropriate
scaling relationships.  In the only published work known to us\footnote{Padoan et al.
1999 have recently submitted a comparison of the Padoan \& Nordlund (1999)
simulations with $^{13}$CO maps of the Perseus Molecular Cloud to the {\it Astrophysical
Journal}.  The cubes are compared using moments of the distribution of line parameters
(see \S 3.5).} that specifically evaluates hydrodynamic simulations by comparing them with
real spectral maps, Falgarone et al. (1994) have compared a simulation by Porter, Pouquet
\& Woodward (1994) with an observed data cube (See also the analysis of simulated cubes in
Dubinski, Narayan
\& Phillips 1995).  The observed cube is a CO map of a small piece of
the expanding H I loop in Ursa Major, first mapped by Heiles (1976).  The Falgarone
et al. (1994) analysis is
based on comparing combinations of the moments of the the derived distributions of spectral
line parameters for each cube.  They find that the
moment analysis on the observed maps agrees well with one performed on the
simulations.  We show, below, however, that this comparison may not have
been strict enough, in that the distribution of the SCF for the Porter et
al.  (1994) simulation differs significantly from the distribution calculated
for the observed Ursa Major data cube.

\section{The SCF Algorithm}

The SCF project was developed in order to probe the nature of correlation
in spectral-line maps of molecular clouds.  Unlike other probes like
Scalo's (1984) Autocovariance Function (ACF) and Structure Function (SF),
the SCF is specifically designed to preserve detailed spatial information
in spectral-line data cubes.  The motivations and mathematical background
of the project are discussed in Goodman (1997).

\subsection{The Development of the SCF}

The SCF algorithm centers around quantifying the differences between
spectra.  To begin, a deviation function, $D$, is defined which represents
the differences between two spectra, $T_1(v)$ and $T_0(v)$.

\begin{equation}
D(T_1,T_0) \equiv \min_{s,\ell}\left\{\int \left[ s \cdot T_1(v-\ell) -
T_0(v) \right]^2 dv \right\}
\end{equation} 
The two parameters $s$ and $\ell$ are included in the
function so differences in height and velocity offset between the
two spectra can be eliminated, recognizing similarities solely in the
shape of two line profiles.  These parameters can be adjusted in order
to find the scaling and/or velocity-space shifting which minimizes
the differences between the spectra. In addition, the deviation
function can be evaluated with either or both of the parameters fixed.

We normalize the deviation function to the unit interval:  a value of 1
indicating identical spectra and a value of 0 indicating minimal
correlation\footnote{A value of 1 can only be achieved in the case of
infinite signal to noise.  See \S \ref{noise}}.  The appropriate
normalization is to divide by the maximum value of the deviation function
in the absence of absorption and subtract this value from 1.  The resulting
function is referred to as the SCF evaluated for the two spectra.

\begin{equation}
S(T_1,T_0) \equiv 1-\sqrt{\frac{D(T_1,T_0)}{s^2 \int T_1^2(v) dv 
+ \int T_0^2(v) dv}}
\end{equation}
As mentioned previously, the deviation function can be evaluated with the
parameters $s$ or $\ell$ fixed, to 1 and/or 0, respectively.  Such restrictions provide
different kinds of information about the two spectra under examination.  The resulting
forms of the spectral correlation function are summarized in Table
\ref{table1}.

In order to examine spectral-line maps, comparison of two spectra must be
extended to the analysis of many spectra simultaneously.  The simplest such
extension is to evaluate the functions $S$, $S^\ell$, $S^s$ and $S^0$
between a base spectrum and each spectrum in the map within a specified
angular range from the base spectrum.  We refer to the angular range under
consideration as the resolution of the SCF.  All of the SCF calculations are
performed using only the central portion of the spectra, specifically, over a range
equal to a given number of FWHMs of the base spectrum from each spectrum's velocity
centroid.  The FWHM is defined by a Gaussian fit to the base spectrum's line
profile.\footnote{The Gaussian fit is {\it only} used to set a reasonable window
over which the SCF is calculated.  For very noisy data, including extra baseline
decreases the SCF. The line profiles discussed in this paper are all roughly
gaussian, with widths that do not vary much within a map, so the change in the SCF
caused by varying window size is tiny.  However, in other cases, such as analysis of
H I line profiles, a window must be manually set, and fixed, as Gaussian fitting
gives spurious widths.  A new version of the SCF, developed to deal with H I data,
uses a fixed spectral window and is presented in Ballesteros-Paredes,
Vazquez-Semadeni \& Goodman 1999.}  The resulting values of the SCF are then
averaged together with the option of weighting the results based on distance from
the original spectrum.  The averaged value is the correlation of the base spectrum
with its neighbors and a similar analysis is then performed for every spectrum in
the map.  When the SCF is evaluated for neighboring points in the spectral-line map,
the averages use many of the same spectra, implying that SCF values for points
within an SCF resolution element are not independent. 

\subsection{The Effects of Instrumental Noise\label{noise}}

Our measure of the similarity between observed spectra must deal with the
effects of noise.  Noise obscures similarities and differences between the
two spectra under examination, usually skewing the results to indicate less
correlation than is actually present.  Hence, the principal difficulty
generated by noise is that it creates a bias in correlation measurements,
favoring spectra with higher values of signal-to-noise.

\begin{figure}
\begin{center}
\plotone{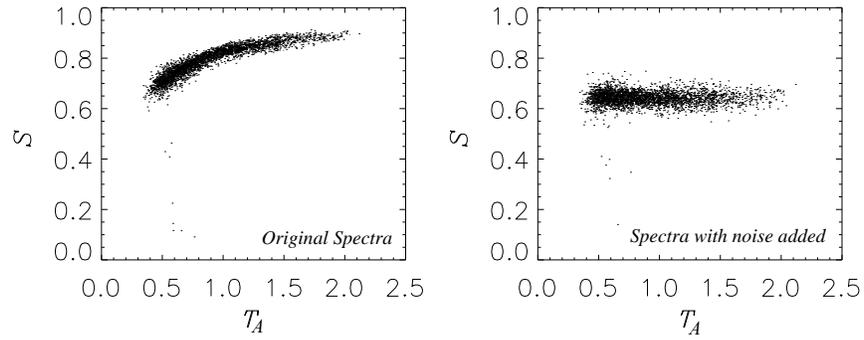}
\caption{\label{figure1} SCF values as a function of antenna temperature 
for (a) the original data with a roughly constant noise but varying $T_A$
and (b) noise added to the spectra to create a uniform
$T_A/\sigma$ ratio of 5.  The bias in correlation for higher values of 
$T_A$ disappears with the addition of noise.  The Heiles Cloud 2 data set
\protect\cite{devries} was used to produce these data.}
\end{center}
\end{figure}

\begin{figure}
\begin{center}
\plotone{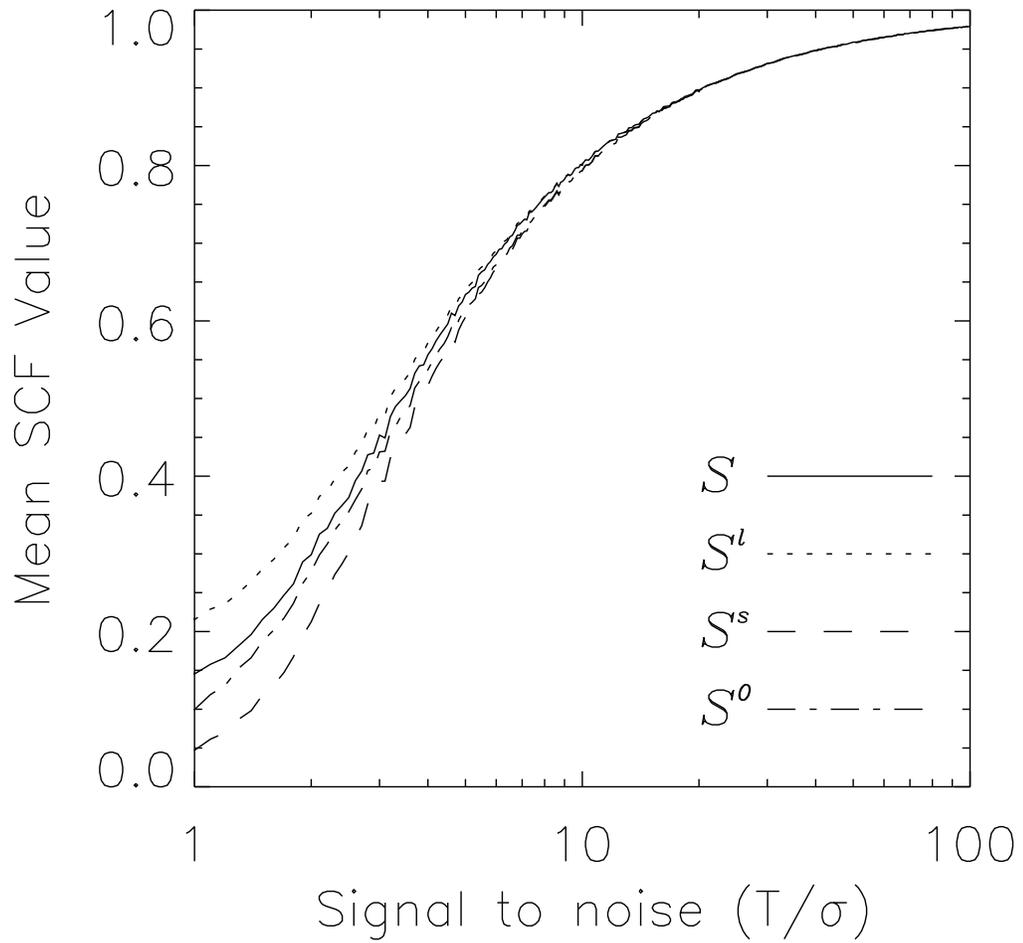}
\caption{\label{figure2} The behavior of the mean values of the
correlation functions with changing signal-to-noise on a cube of 
Gaussian spectra.}
\end{center}
\end{figure}

We have explored a few methods of subtracting out noise bias using
techniques shown to work on infinitely well-sampled data.  However, these
techniques break down for data with limited resolution \cite{roz}.  While
highly unconventional, we have found that the best method of dealing with
non-uniform signal-to-noise is to discard all spectra with signal-to-noise
below a certain cutoff value, $(T_A/\sigma)_c$, and then to add normally
distributed random noise to spectra with signal-to-noise greater than the
cutoff until all spectra have $T_A/\sigma=(T_A/\sigma)_c$.  This method
appears effective because it eliminates the bias (See Figure \ref{figure1})
and the resulting correlation outputs do not appear to depend strongly on
specific set of noise added.  The maximum value of the SCF cannot reach 1
for any finite value of $T_A/\sigma$; instead, the maximum is dependent on
the line shape and the cutoff value for the signal-to-noise.  Figure
\ref{figure2} depicts the rise in SCF values with increasing
signal-to-noise for each of the correlation functions.  These data are
generated by using the SCF to analyze a data cube consisting of identical
Gaussian spectra which have had noise added to achieve a specific value of
$(T_A/\sigma)_c$.

We considered renormalizing the SCF by a factor equal to the inverse of the
maximum SCF possible for the $(T_A/\sigma)_c$ used.  If we did so then
absolute SCF values would always have the same meaning.  However, since the
exact maximum possible depends on line shape, we chose not to renormalize.
Instead, we note that whenever $(T_A/\sigma)_c$ is set to the same value,
the maximum possible SCF value should be roughly equal for any cube.  In
the examples below, we set $(T_A/\sigma)_c=5$, which implies a maximum
possible SCF of order 0.65 (See Figure \ref{figure2}).

This ``noise equalizing" procedure may seem distasteful to some--especially to
observers who spend long hours at the telescope!   The best way we can explain its
necessity is by reminding the reader that you cannot get something for nothing. 
In other words, if your data is noisy, you simply {\it cannot} know how well
correlated two spectra are as well as if the spectra were clean.  Uncorrelated noisy spectra can look just the same as correlated noisy
spectra. Any correction introduces different amounts of uncertainty for different positions
in the map.   For this reason,
until someone makes a better proposal, we continue to advocate equalizing signal to noise
at a threshold value.

\section{First Results from the SCF}

In this section, we analyze five sample data cubes
chosen to demonstrate the SCF's ability to discriminate among different physical
conditions.  First, we describe the data cubes (see Table 2), and then we discuss
comparisons amongst them in the context of the SCF.  Of the two observational data cubes,
one is for a self-gravitating cloud (Heiles Cloud 2), and one is for a
non-self-gravitating high-latitude cloud (in Ursa Major).  For the three cubes generated
from simulations: one is purely hydrodynamic and non-self-gravitating; one is
magnetohydrodynamic and non-self-gravitating; and the last is magnetohydrodynamic {\it
and} self-gravitating. 

\subsection{Data Sets}

\noindent
{\it Heiles Cloud 2:}  In 1996, observers at FCRAO mapped Heiles Cloud 2
in the $\mbox{C}^{18}\mbox{O}(2-1)$ line (deVries et al. 1999).  The resulting data cube
consists of 4800 spectra arranged in a grid of
$50 \times 96$ pixels on the sky.  The grid covers $58^{\prime}$ in right ascension and
$40^{\prime}$ in declination, centered at $\alpha(2000)=4^h 36^m 09^s$ and
$\delta(2000)=25^{\circ}47' 30''$.  Assuming the cloud is 140 pc distant, the
map covers a physical area of $2.3 \times 1.6$ pc at a spatial resolution
of 0.034 pc.  The spectra have 256 channels of velocity running from -0.35
km/s to 12.45 km/s, with a channel width of 0.05 km/s.  The peak emission
from the cloud is at about +6\ km/s.  

\noindent
{\it Ursa Major:}  The $^{12}$CO (2-1) map analyzed here is described in Falgarone
et al. 1994.  The area mapped is located on a giant H I loop, and is claimed by Falgarone
et al. to be ``a good site to study turbulence in molecular clouds given its proximity to
an important source of kinetic energy (the expanding loop itself)."  The size of this map
is 9 by 19 pixels, with a grid step of 30$''$ (0.015 pc if the cloud is at 100 pc).  Note
that there are approximately 50\% more pixels in the Porter et al. simulation
than in this relatively small map.

\noindent
{\it Pure Hydrodynamic Turbulence:}
This cube, presented in  presented in Porter et al. (1994), is the one compared with the
$^{12}$CO (2-1) Ursa Major data cube (see above) by Falgarone et al. (1994).  It is a
three-dimensional simulation with periodic boundary conditions, no magnetic fields,
no gravity, and fully compressible turbulence.  The time step analyzed
here is the second cube  ($t=1.2\tau_{ac}$) presented in Falgarone et al. (1994).

The spectra in the simulated data cube are laid out in a grid of $16 \times
16$ pixels.\footnote{The full Porter et al. 1994 simulation is 512$^3$, but the grid of
$16 \times 16$ spectra is generated by considering subsamples of the cube with dimensions
$32
\times 32
\times 512$.}  There are 512 channels in each spectrum and the channel width is 0.13
km/s.  The simulated spectra are generated from density-weighted histograms of velocity,
which are intended to mimic observations of the
$^{12}$CO(2-1) line.  The overall physical size of the simulation is not given, but the
nature of comparison with the CO map of Ursa Major shown in Falgarone et al. (1994)
implies that the spatial resolution of the simulated spectral-line map should be
approximately 0.015 pc (30$''$ at 100 pc).  

\noindent
{\it Magnetohydrodynamic Turbulence:} Mac Low et al. (1998) have made their
simulated cube ``L," which represents uniform, isotropic, isothermal, supersonic,
super-AlfvŽnic, decaying turbulence available to us and others through the world wide
web.  The spectra are produced as density-weighted histograms from a simulation using a
finite-difference method (ZEUS; see Stone \& Norman 1992) and 256$^3$ zones.  Mac Low et
al. use this cube, and others, to study the free decay of turbulence in the ISM.    
Mac Low et al.'s  results concerning decay times apparently agree with those of Padoan \&
Nordlund (1999), who have carried out equivalent simulations.   The
physical scale the Mac Low et al. cube represents depends on the choice of other
paramenters (e.g. field strength), but it is fair to estimate that the resolution should
be similar to that of Gammie et al (1999; see below), or about 0.06 pc.

\noindent
{\it Self-Gravitating Magnetohydrodynamic Turbulence:}  Another group, whose most
recently published work is Stone, Ostriker, \& Gammie (1998), has been using the ZEUS
code to study self-gravitating MHD turbulence.  Charles Gammie has kindly provided us with
a preliminary simulated spectral-line cube with dimensions $32 \times 32 \times 256$,
generated from a recent 3D, self-gravitating, high-resolution run (Gammie et al. 1999). 
The spectra are density-weighted histograms meant to simulate $^{13}$CO emission, observed
with a velocity resolution of 0.054 km s$^{-1}$.   As is the case for the Porter et
al. (1994) simulations presented in Falgarone et al. (1994), the larger original (here,
256$^3$) simulation is downsampled in the two spatial dimensions (here to $32 \times 32$)
to produce reasonable spectra.  The resulting spatial resolution is approximately 0.06 pc. 

\subsection{Analysis}

For all of the SCF analyses presented in this paper, the cutoff
signal-to-noise ratio is set to 5, the spatial resolution of the SCF
includes all spectra within 2 pixels of the base spectrum, uniform
weighting is applied across the resolution element, and the portion of each
spectrum within 3 base spectrum FWHMs of the velocity centroid is used.

\begin{figure}
\plotone{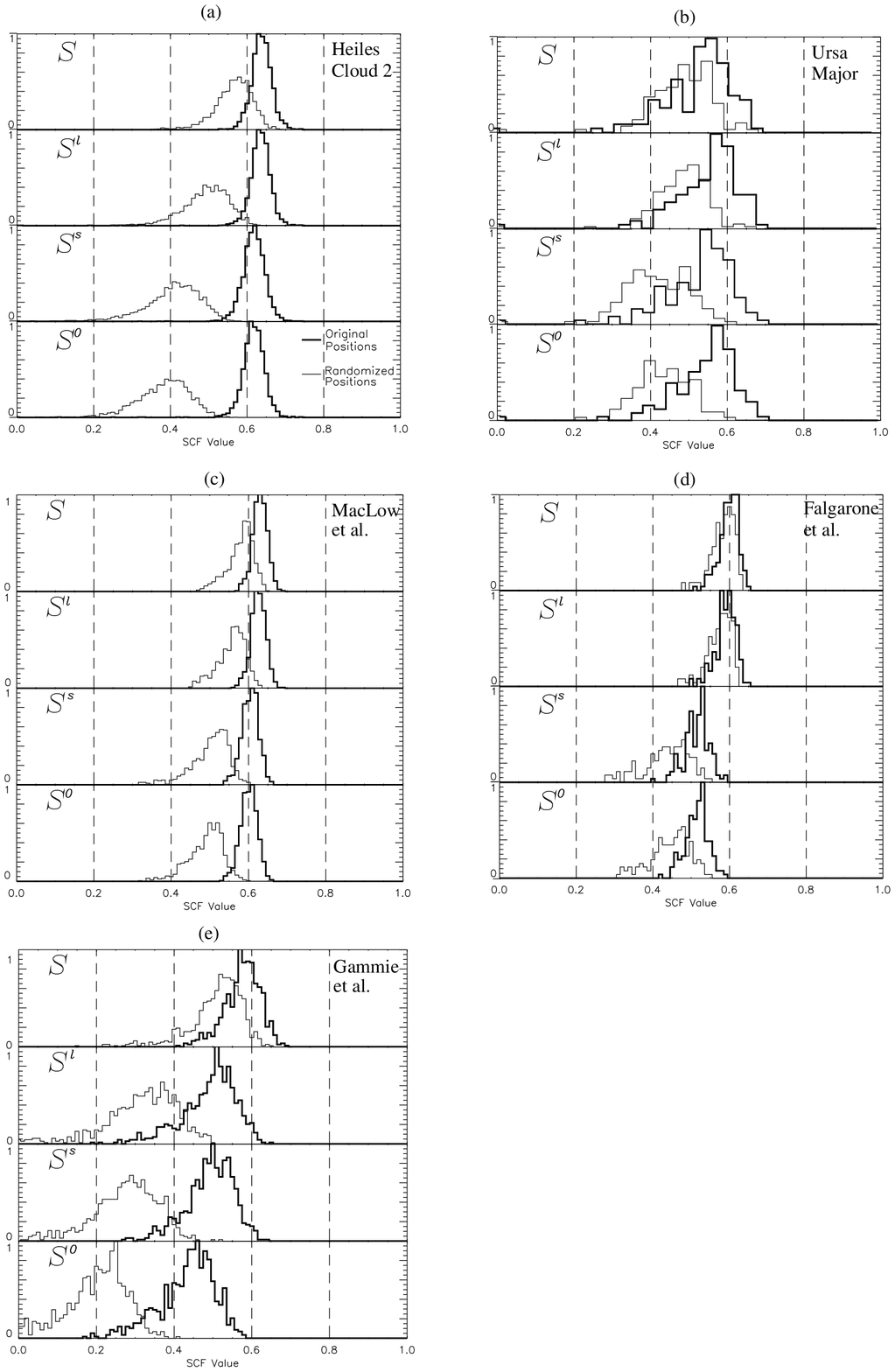}
\caption{\label{figure3} Figure 3 caption on next page}
\end{figure}

For each data cube, greyscale maps of the SCF values are generated and compared with
maps of line parameters such as antenna temperature.  Note that before the
noise correction discussed in \S\ref{noise} was applied, maps of the peak
antenna temperature $T_A$ and the SCF looked similar.  After correction,
this is no longer the case.  So, the greyscale maps of the SCF, which preserve all the
spatial information about which spectra in a map are correlated, can be informative on
their own.  In addition to pointing out correlations with various line parameters, they are
highlight edges of H I shells (see Ballesteros-Paredes, Vazquez-Semadeni \&
Goodman 1999) and other such structures.  In the current paper, however, we
show only simple histograms (Figure \ref{figure3}) of all of the values of the SCF in a
map, which are easier to quantitatiely intercompare than the greyscale images.   We
present a moment analysis of these distributions in Table 3.

As a test of the hypothesis that the positions of spectra within a cube are
important to its description, we calculate the SCF for the original cubes
and for ``comparison'' cubes where the positions are randomized.  If the
meaning of the SCF is linked to the original positions of the spectra,
randomization of the positions should create a significant change in the
SCF values.  This drop is in fact observed in our analysis.  The magnitude
of the drop depends on the cube being analyzed and the compensatory
parameters used in the SCF but not on the specific randomized positions.  Different
randomizations produce changes of
$\sim 1$\% in the mean values of correlation functions.  The SCF results for the
randomized cubes appear with the original histograms and moment analysis (Figure
\ref{figure3} and Table 3).

The significance of the drop in the mean value of the SCF caused by randomization is
high in all cases studied here.  The error in estimating the mean in each SCF
distribution is of order the standard deviation of the distribution divided by
$\sqrt{N_{\rm pixels}}\times 3 \Delta v_{\rm channels}$, where $N_{\rm pixels}$ is
the number of pixels in the map and $3\Delta v_{\rm channels}$ is the number of
spectral channels considered in calculating the SCF.  As can be seen from Tables 2
and 3, the standard deviations of the distributions are all less than 0.1, and the
minimum $\sqrt{N_{\rm pixels}}\times 3 \Delta v_{\rm channels}$ is of order $5
\times 10^3$, meaning that the error in the quoted mean is always less than about $1
\times 10^{-3}$.  This value is much smaller than any of the differences between SCF
means for actual and randomized maps listed in the last column of Table 3.

\subsection{Inferences}

The most basic conclusion we draw from our analysis is that the SCF does
recognize some form of spectral correlation in data cubes.  The drop in the mean and
increased spread in the distribution when positions are randomized depicts a
clear loss of spatial correlation of the spectra.  When the compensatory
parameters $s$ and $\ell$ are fixed, the difference between the two
histograms becomes clearer because the program has fewer tools to
compensate for the differences in the spectra.  For example, in Heiles Cloud 2,
randomization causes the smallest correlation drop when both the lag and
scaling parameters are allowed to vary, indicating that the spectra are similar
in shape throughout the data cube.  The fact that the mean value 
of $S^{\ell}$ is
larger than that of $S^s$ implies that the spectra 
in the cube are more similar in
overall antenna temperature than in velocity distribution.  In other words,
it appears that the compensatory parameter $\ell$ is more important for
good correlation than is the scaling parameter $s$ in this case.

Examining Figure \ref{figure3} and Table 3 closely, one can detect a clear pattern. 
From this set of examples, it appears that {\it the more gravity matters, the more of an
effect randomization has on the SCF distribution.} The SCF distribution for the
self-gravitating cloud (Heiles Cloud 2) shows a much larger change in response to
randomization in spectral positions than does the unbound high-latitude cloud (Ursa
Major).  The simulation which most closely reflects the Heiles Cloud 2 response to
randomization is the only one that includes self-gravity: Gammie et al. (1999).  In the
non-self-gravitating case, the Ursa Major SCF distributions show much less change in
response to randomization than the Heiles Cloud 2 distributions, but more change than the
distributions for the simulations presented as their ``match" in Falgarone et al. 1994
(see Table 3). This trend is corroborated by a visual inspection of the grid of simulated
spectra (Falgarone et al. 1994 Figure 1b), where it is seen that the differences between
neighboring spectra appear more pronounced than in the Ursa Major observations. Such
behavior indicates that the original positions of the spectra in the simulated cube are
{\it not} as essential to the cloud description as they are in Ursa Major.    

As for the mean values of the SCF, it is acceptable to intercompare means for the five data
sets used here because signal-to-noise values have been equalized (see \S \ref{noise}). 
The SCF means for the Mac Low et al. (1998) simulations seem the best match to the Heiles
Cloud 2 data.  The means for the simulation presented in Falgarone et al. 1994 are not
very similar to those for the Ursa Major cloud which Falgarone et al. claim is an excellent
match for these simulations.  In particular, the means for $S$ and $S^\ell$ in the
simulation are nearly equal (0.6) as are those for $S^s$ and $S^0$ (0.52), meaning that lag
adjustments are important, but scale adjustments are not.  In the Ursa Major
observations, all forms of the SCF give roughly 0.55, and this difference between lag and
scaling is not seen.

\subsection{Resolution and Sampling Effects}
If two maps of similar regions have very different spatial resolution, the map with lower
resolution will look smoother.  One can imagine that the spectra in the lower
resolution map will change more rapidly from one position to the next, so that the mean
SCF for the lower resolution map will be lower if the SCF is run with the same size
averaging box.  Alternatively, one can imagine running the SCF with a very large averaging
box on a high-resolution map, thus ``smearing out" small differences in spectra that might
show up in a smaller box. 

To investigate issues of resolution, we have used the Heiles Cloud 2 data set in a
numerical experiment.  We re-ran the SCF many times, using a box size of 3 (the original),
5, 7, 9, 11, 13, and 15 pixels, to see what effect this ``smoothing" would have on the
SCF.  Figure 4 presents the results of this experiment.

The mean value of the SCF in Heiles Cloud 2 does drop for larger box sizes, but by a
surprisingly small amount (see Figure 4).  (For the Heiles Cloud 2 cube with positions
randomized, the SCF is unaffected by changes in resolution, as expected.)  The width
of the SCF distributions for Heiles Cloud 2 with positions randomized also drops a
bit for larger box sizes, illustrating that small spectral differences do eventually
get smeared-out, even when spectra are shuffled.  Thus, for this one example, the
effect of changing resolution is relatively small.

Other factors can also effect the true resolution--and resulting magnitude--of the
SCF. The way the spectra in a map are sampled will effect the absolute values of the
SCF measured.  For example, in a Nyquist-sampled map, neighboring spectra are not
independent and will necessarily yield higher values of the SCF than a beam-sampled
map.  In this paper, we have tried to restrict ourselves to beam-sampled maps, but a
correction for sampling should be added to the SCF algorithm in the future.
 
Ultimately, it is the relationship between map resolution, averaging box size, map size,
and physically important scales (e.g. Alfv\'en wave cutoff, outer scale of turbulence,
Jeans length, etc.) that determine the SCF.  The subtleties of these relationships'
influence on the SCF, and on other statistical techniques, are discussed in Padoan \&
Goodman 1999.

\subsection{How discriminating is the SCF?}
This paper is intended  as a ``proof-of-concept," to show that the SCF is a 
discriminating tool for analyzing observed and simulated spectral-line maps.
As discussed in the previous section, Falgarone et al. (1994) concluded that the
simulated cube of Porter et al. 1994 is an extraordinary match to the $^{12}$CO(2-1)
observations of Ursa Major.  This conclusion is based on Falgarone et al.'s analysis of
combinations of statistical moments of distributions of antenna temperatures.  We have
shown here that, contrary to Falgarone et al.'s conclusions, the SCF can detect significant
{\it differences} between these two data sets.  

In addition, we have tested the SCF against the comparison method where the
simple histograms of moments (centroid velocity, velocity width, skewness, and
kurtosis) of spectra in a map are used to intercompare data cubes.  It turns out that this
seemingly simple method has one major weakness.  The values of the higher-order moments
(skewness and kurtosis) are extremely sensitive to how one treats noise in the data cube. 
Some researchers (e.g. Padoan et al. 1999) choose to set a threshold of $n\sigma$ (where
$n$ is usually 3 and $\sigma$ is the rms noise in a spectrum) before calculating moments. 
It turns out that the value of $n$ has profound effects on the skewness and kurtosis
distributions for the whole cube.  We also compared equalizing the signal-to-noise in a
cube and then computing moments.  This gives {\it different} results than thresholding.  
Given these complications, we reserve further discussion of moment distribution
comparisons for future work.

For now, based on the re-comparison of the data and simulation in Falgarone et al. 1994,
we have shown that the SCF {\it does} find differences where other methods do not.

\section{Conclusions}

The drop to lower correlation values when spectral positions are
randomized shows the SCF is performing its primary function
of quantifying the correlation between proximate spectra in a data cube.

In this first paper, we have demonstrated that the SCF algorithm can
find subtle differences between simulated and observational
data cubes that may not be evident in other kinds of comparisons.  
Thus, application of the SCF will provide a ``sharper tool'' to be
used in comparing simulated and observational data cubes in the future.
There is far more information produced by the SCF algorithm than is
presented here and the full extent and import of the information
provided will be explored in subsequent papers.

In the future, we plan to exploit the information generated by the
SCF to examine a large variety of observational and simulated data cubes.
Ultimately, our aim is to use the SCF to evaluate which physical conditions
imposed on MHD simulations are necessary to produce the correlations
observed in the ISM.  The initial results presented here, showing stronger drops in
spectral correlation in response to randomization in self-gravitating situations,  
hint that including self-gravity may be {\it essential} in the numerical recreation of
star-forming regions. Physical effects other than gravity, such as large-scale shocks,
should also have an ``organizing" effect on the spatial distribution of spectra, and we
intend to search for those effects with the SCF as well.

A deeper coverage of the development of the SCF is
found in Rosolowsky (1998) which is available on the internet\footnote{See
http://cfa-www.harvard.edu/\~{}agoodman/scf/SCF/scfmain.html}.  The SCF
algorithm is written in IDL and is available for public use\footnote{
http://cfa-www.harvard.edu/\~{}agoodman/scf/distribution/ }.

\acknowledgements

We would like to thank Marc Heyer of FCRAO for the use of the Heiles Cloud
2 data set prior to publication.  We wish to express our thanks to Derek
Lis who facilitated access to the simulated and observed data in Falgarone et
al.  (1994).  Uros Seljak provided some clear insights into the effects of
instrumental noise which have proved indispensable in our understanding and
we are grateful.  Thanks are in order to Mordecai Mark Mac-Low for
providing access to his group's MHD simulations.  And, our gratitude
extends to Charles Gammie who allowed access to the MHD simulations of
Gammie, Ostriker and Stone.  Analysis of these additional simulations is available
at our web site.  Javier Ballesteros-Paredes provided valuable help with
revisions to this paper.  Finally, we are most grateful to an anonymous referee
whose comments very significantly improved the integrity of this paper.  This
research is funded by grant AST-9721455 to A.G. from the National Science
Foundation.
\clearpage

\begin{table*}
\begin{center} 
\caption{\label{table1} Summary of Correlation Functions}
\begin{tabular}{cccl} 
\hline 
Function Name & $s$ & $\ell$ & Spectral Property Examined \\
\hline
$S$ & Float & Float & Compares shapes of spectra \\
$S^{\ell}$ & 1 & Float 
& Emphasizes similarity in shape and amplitude \\
$S^s$ & Float & 0 
& Highlights similarity in velocity offset and shape \\
$S^0$ & 1 & 0 & Measures similarity in all properties \\
\hline
\end{tabular} 
\end{center}
\end{table*}
\clearpage
\vskip 2in
\begin{table}
\caption{\label{table2} Summary of Data Cubes Analyzed} 
\plotone{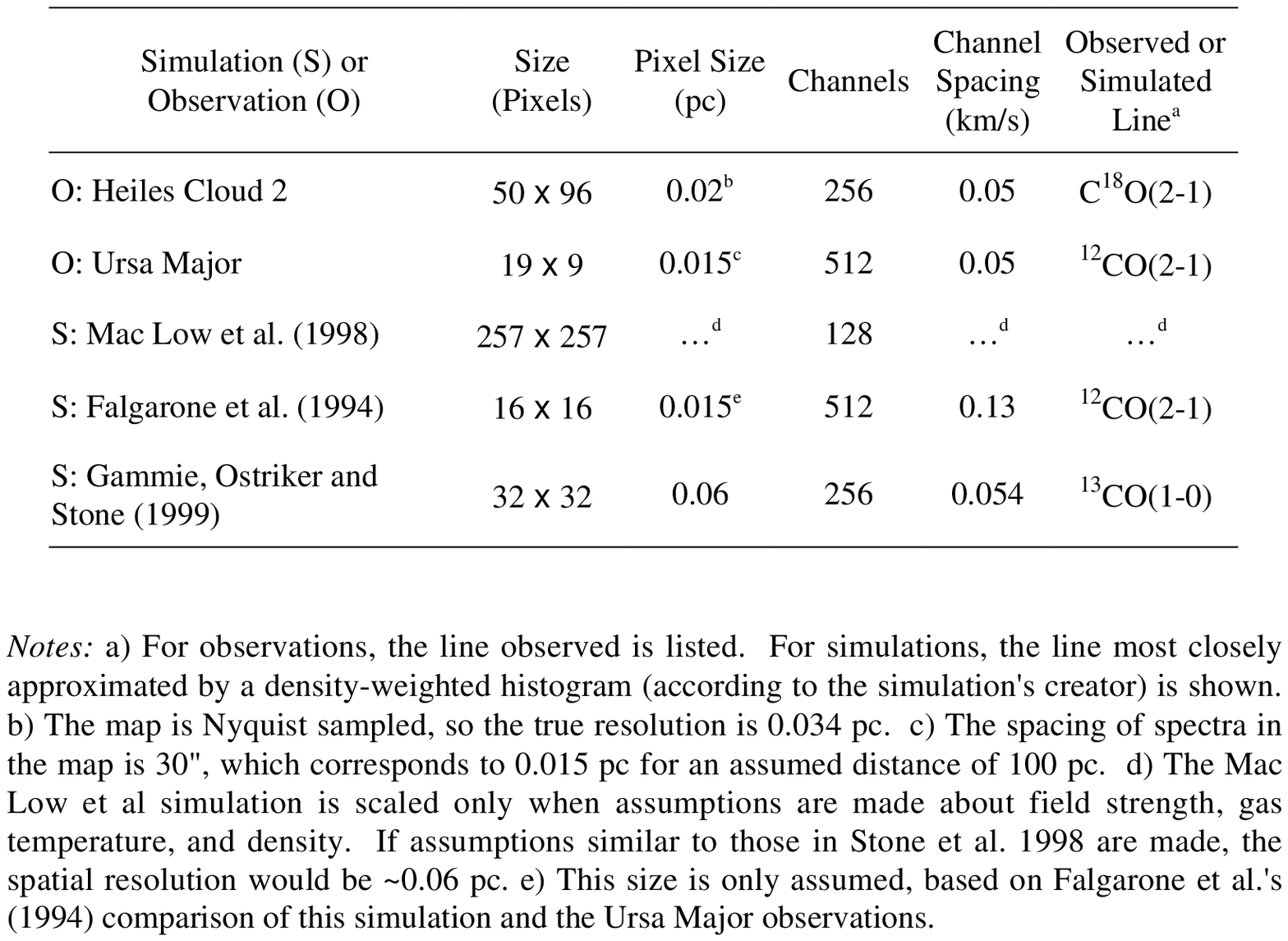}
\end{table}
\vskip .2in
\clearpage
\begin{table}
\caption{\label{table3} Statistical Outputs of the SCF} 
\begin{center}
\begin{tabular}{cccccc}
\tableline\tableline
& \multicolumn{5}{c}{Heiles Cloud 2 (Star-forming Cloud)}\\
\cline{2-6}
Function & \multicolumn{2}{c}{Original Position} & 
\multicolumn{2}{c}{Random Positions}&\raisebox{-1ex}[0pt]{Difference}\\
\cline{2-5}
& Mean & Std. Deviation & Mean & Std. Deviation &
\raisebox{1ex}[0pt]{in Means} \\
\tableline
$S$ &0.64 & 0.030 & 0.57 & 0.051 & 0.07 \\
$S^{\ell}$ & 0.63 & 0.033 & 0.50 & 0.061 & 0.14 \\
$S^s$ & 0.62 & 0.037 & 0.41 & 0.074 & 0.21\\
$S^0$ & 0.62 & 0.036 & 0.39 & 0.070 & 0.23 \\
\tableline
&\multicolumn{5}{c}{Ursa Major (Unbound High-Latitude Cloud)} \\
\cline{2-6}
Function & \multicolumn{2}{c}{Original Position} & 
\multicolumn{2}{c}{Randomized Positions}&\raisebox{-1ex}[0pt]{Difference}\\
\cline{2-5}
& Mean & Std. Deviation & Mean & Std. Deviation &
\raisebox{1ex}[0pt]{in Means} \\
\tableline
$S$ & 0.55 &0.10  &0.49  &0.09  &0.05  \\
$S^{\ell}$ & 0.56 &0.09  &0.49  &0.08  &0.07  \\
$S^s$ & 0.54 &0.10  &0.42  &0.09 &0.12  \\
$S^0$ & 0.55 &0.09  &0.44  &0.08  &0.11  \\
\tableline
\end{tabular}
\end{center}
\end{table}

\clearpage

\begin{table}
%\caption{\label{table3a} Statistical Outputs of the SCF (continued)} 
\begin{center}
\begin{tabular}{cccccc}
\tableline\tableline
& \multicolumn{5}{c}{Mac Low et al. 1998 (MHD, no gravity)}\\
\cline{2-6}
Function & \multicolumn{2}{c}{Original Position} & 
\multicolumn{2}{c}{Random Positions}&\raisebox{-1ex}[0pt]{Difference}\\
\cline{2-5}
& Mean & Std. Deviation & Mean & Std. Deviation &
\raisebox{1ex}[0pt]{in Means} \\
\tableline
$S$ &0.64 &0.02  & 0.58 & 0.03 &  0.06\\
$S^{\ell}$ & 0.63 & 0.02 &  0.56& 0.04 & 0.07 \\
$S^s$ & 0.61 & 0.02 &0.51  & 0.05&0.10 \\
$S^0$ &0.61  & 0.02 &0.50  & 0.05 & 0.11 \\
\tableline
&\multicolumn{5}{c}{Falgarone et al. 1994 (HD, no gravity)} \\
\cline{2-6}
Function & \multicolumn{2}{c}{Original Position} & 
\multicolumn{2}{c}{Randomized Positions}&\raisebox{-1ex}[0pt]{Difference}\\
\cline{2-5}
& Mean & Std. Deviation & Mean & Std. Deviation &
\raisebox{1ex}[0pt]{in Means} \\
\tableline
$S$ & 0.60 & 0.020 & 0.59 & 0.030 & 0.01 \\
$S^{\ell}$ & 0.59 & 0.026 & 0.58 & 0.033 & 0.02 \\
$S^s$ & 0.52 & 0.032 & 0.45 & 0.060 & 0.07 \\
$S^0$ & 0.52 & 0.030 & 0.45 & 0.052 & 0.07 \\
\tableline
&\multicolumn{5}{c}{Gammie, Stone \& Ostriker 1997 (MHD, with gravity)} \\
\cline{2-6}
Function & \multicolumn{2}{c}{Original Position} & 
\multicolumn{2}{c}{Randomized Positions}&\raisebox{-1ex}[0pt]{Difference}\\
\cline{2-5}
& Mean & Std. Deviation & Mean & Std. Deviation &
\raisebox{1ex}[0pt]{in Means} \\
\tableline
$S$ & 0.58 & 0.05 & 0.52 & 0.07 & 0.06 \\
$S^{\ell}$ & 0.50& 0.07 & 0.32 & 0.09 & 0.18 \\
$S^s$ & 0.50 & 0.06 & 0.28 & 0.09 & 0.22 \\
$S^0$ & 0.44 & 0.07 & 0.21 & 0.08 & 0.23 \\
\tableline\tableline
\end{tabular}

\end{center}
\end{table}

\clearpage

\noindent
Figure 3--Histograms of SCF values for observed
and simulated data cubes.
To facilitate comparison between the correlation functions, histograms have been normalized
to the unit interval for the unrandomized cube, and to an integral equal to the
unrandomized cube for the randomized cube. The histogram shown in heavy print represents
the correlations for the spectra in their original positions and the lighter line
indicates the distribution for randomized positions. Distributions are shown for: (a)
observed C$^{18}$O map of the star-forming cloud Heiles Cloud 2
\protect\cite{devries}; (b) observed $^{12}$CO(2-1) map of the Ursa Major unbound
high-latitude cloud (Falgarone et al. 1994); (c) magnetic, non-self-gravitating
simulation of Mac Low et al. 1998; (d) non-magnetic, non-self-gravitating simulation of
Porter et al. \protect (1994) cube used by Falgarone et al. \protect (1994); and (e)
magnetic, self-gravitating simulation of Gammie et al. (1999). Table 2 lists the properties
of the data sets illustrated, and Table 3 compares the means and standard deviations of
distributions shown here.  The four variants of the SCF are described in Table
\ref{table1}.

\clearpage
\begin{figure}
\begin{center}
\plotone{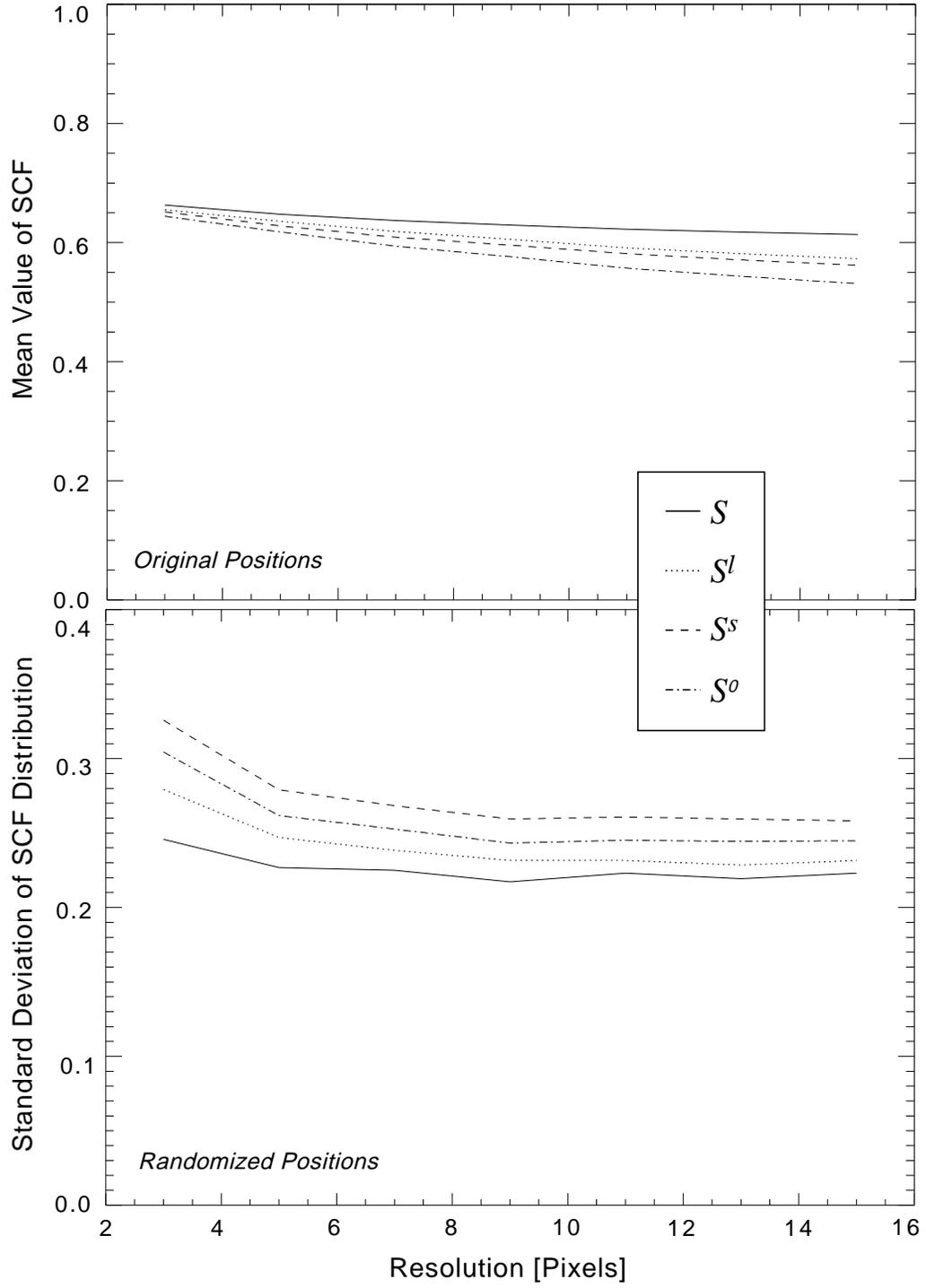}
\caption{\label{figure4} Figure 4 caption on next page.}
\end{center}
\end{figure}
\clearpage
\noindent
Figure 4 -- The behavior of the SCF as a
function of changing resolution.  For the HCl2 data set with uniform signal-to-noise
$(T_A/\sigma)_c= 5$, the top panel shows the mean value of the SCF as a function of the
size of the box over which the SCF is calculated. The bottom panel shows the $1-\sigma$
width of the distribution of the SCF, for the same noise-equalized HCl 2 data set,
but with positions randomized.  For any randomized cube, the mean of the SCF is
{\it independent} of resolution, and only the width of the distribution changes,
as shown.  In creating these plots, runs using 3, 5, 7, 11, 13, and 15 pixel square
sampling areas for the SCF are used.
\end{document}